\documentclass[aps,notitlepage,nofootinbib]{revtex4-1}

\usepackage{bm,amsmath,amssymb,graphicx,MnSymbol}

\usepackage{footnote}

\usepackage{enumitem}

\newcommand{\uvc}[1]{\bm{\mathrm{\hat #1}}} 

\newcommand{\bx}{{\bm x}}
\newcommand{\ba}{{\bm a}}

\usepackage[format=plain,labelfont={bf,small},textfont=small,justification=raggedright,singlelinecheck=false]{caption}

\usepackage{tikz}

\begin{document}

\title{Momentum and pseudomomentum in a shallow water equation}
\author{J. A. Hanna}
\email{jhanna@unr.edu}
\affiliation{Mechanical Engineering, University of Nevada, 1664 N. Virginia St. (0312), Reno, NV 89557-0312, U.S.A.}

\date{\today}

\begin{abstract}
A basic shallow water system with variable topography is analyzed from the point of view of a Lagrangian derivation of momentum, energy, and pseudomomentum balances.  A two-dimensional action and associated momentum equation are derived.  The latter is further manipulated to derive additional equations for energy and pseudomomentum.  This revealed structure emphasizes broken symmetries in space and a reference configuration, and preserved symmetry in time. 
\end{abstract}

\maketitle

In a recent paper, Singh and the present author \cite{SinghHanna21,*SinghHanna21correction} explored a general framework for the balance laws of continuous mechanical systems described by an action, with emphasis on preserved or broken symmetries in space, time, and, particularly, the material continuum itself.
In one example, this approach provided useful insights into the case of a non-uniform elastic rod moving through a non-uniform environment, a system in which broken material and spatial symmetries lead to distinct source terms in the corresponding balances of pseudomomentum and momentum.
The present note examines another, entirely different, example of such a system, namely the motion of shallow water on variable topography.  The topography provides two things: an initial condition that appears as a non-uniform property of an effectively two-dimensional thin layer of fluid, and a non-uniform background field through which this layer subsequently moves.

Although it seems likely that a close relationship exists between the two concepts, the term ``pseudomomentum'' does not here denote a property of waves or other disturbances superimposed on a background flow, as in some geophysical fluid dynamics literature \cite{AndrewsMcIntyre78, Shepherd90, VallisBOOK, BuehlerBOOK}. 
These and related works also derive their results using techniques that differ considerably from the present one.
The present approach will be Lagrangian rather than Hamiltonian, Lagrangian rather than Eulerian, and will not involve a non-quiescent base state or any time-averaging thereof.  
Rotation and associated effects will not be included in the shallow water model. 
The purposes of this brief exercise are to further illustrate the application of the formalism of \cite{SinghHanna21,*SinghHanna21correction}, use its perspective to offer an interpretation of a rudimentary shallow water equation, and hopefully gain preliminary insight towards a possible unification of the two
approaches.

\section{A simple shallow water action}

Several variational formulations of inviscid shallow water equations may be found in the literature \cite{Whitham65-2, Whitham67, Luke67, Miles77, MilesSalmon85, DellarSalmon05, Camassa96, ClamondDutykh12}.  Here we will construct a basic model from first principles.

\begin{figure}[h!]
	\includegraphics[width=5in]{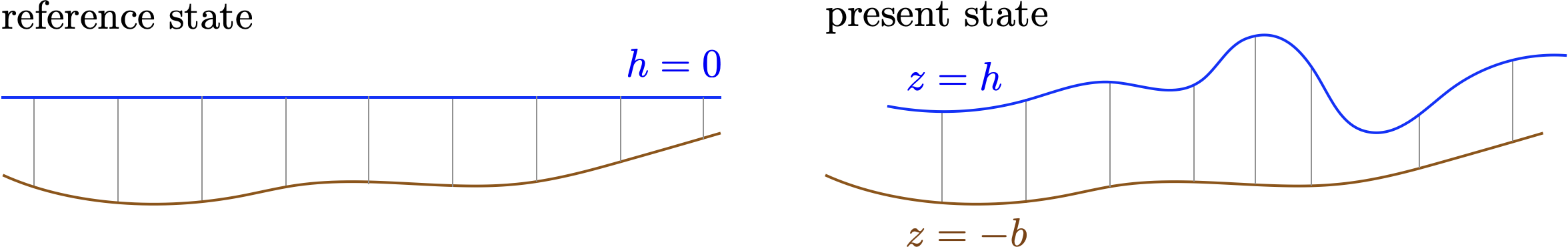}
	\captionsetup{margin=1in}
	\caption{Schematic of reference and present configurations of the fluid layer with free surface $h$ and bottom topography $b$.  The vertical lines represent a possible choice of material markers for one of the planar coordinates.  The Cartesian $z$ is vertical.  Columns of fluid are stretched or compressed vertically in response to waves or topography, preserving three-dimensional volume.  All slopes are shown exaggerated; they are restricted to be small.}
	\label{shallowfig}
\end{figure}

Consider a thin layer of fluid (Figure \ref{shallowfig}) acted on by gravity parallel to the $\uvc{z}$ direction, its free and bottom surfaces given by the functions $z=h$ and $z=-b$, respectively.
The position of a fluid particle labeled by material coordinates $\eta^i$ is given by $\bx(\eta^1,\eta^2) + z(\bx, \bar\bx, \eta^3)\uvc{z}$, where $\bx$ describes two-dimensional position in a plane perpendicular to $\uvc{z}$, and $\bar\bx(\eta^1,\eta^2)$ is a reference position in this plane. 
In the reference state, $h=0$ and $b(\bx) = b(\bar\bx) \equiv \bar b(\bar\bx)$.
The material time derivative is written as $d_t$, and material derivatives with respect to a coordinate are $d_i \equiv \frac{d}{d\eta^i}$.  Latin indices will run from 1 to 3, Greek from 1 to 2.
The kinematic assumption is that the velocity field is incompressible and vertically uniform, meaning that as infinitesimal vertical columns of fluid transmit waves and move over topography, they retain their three-dimensional volume, remain vertical, and their vertical deformation is affine stretching or compression.  
We may then define the third coordinate such that $-\bar b(\bar\bx) \le \eta^3 \le 0$ and $z = h(\bx) + \frac{ h(\bx) + b(\bx) } { \bar b (\bar\bx) }\eta^3$.  Thus, $\eta^3=0$ at the free surface $z=h$, and is just a 
 Cartesian coordinate in the reference configuration.
A further geometric assumption is that the slopes of $h$ and $b$ are sufficiently small that any non-planarity of $\bx$ or skewness of coordinate lines can be neglected.  This implies a restriction to long waves over gentle slopes.  

In keeping with the small-slope approximations, the following quantities are to be thought of as planar two-dimensional objects: 
coordinate bases $\bar{\ba}_\alpha = d_\alpha\bar\bx$ and $\ba_\alpha = d_\alpha\bx$ in the reference and present configurations, 
reciprocal bases defined by $\bar{\ba}^\alpha\cdot\bar{\ba}_\beta = \delta^\alpha_\beta$ and $\ba^\alpha\cdot\ba_\beta = \delta^\alpha_\beta$, corresponding covariant derivatives $\bar\nabla_\alpha$ and $\nabla_\alpha$, which commute with the respective bases (curvature is neglected), and coincide with the material derivative when acting on index-free objects, referential and present planar gradients $\bar\nabla() = \bar\nabla_\alpha()\bar{\ba}^\alpha$ and $\nabla() = \nabla_\alpha()\ba^\alpha$ and divergences $\bar\nabla\cdot(\bar{\ba}_\alpha()^\alpha) = \bar\nabla_\alpha()^\alpha$ and $\nabla\cdot(\ba_\alpha()^\alpha) = \nabla_\alpha()^\alpha$.  The two-dimensional Jacobian determinant $J \equiv \sqrt{a/\bar a}$, where $\bar a = \text{det} [ \bar{\ba}_\alpha \cdot \bar{\ba}_\beta ]$ and $a = \text{det} [ \ba_\alpha \cdot \ba_\beta ]$, will appear frequently, and we will often need the Piola identities $\nabla_\alpha\left(J^{-1}\bar{\ba}^\alpha\right)=\bm{0}$ and $\bar\nabla_\alpha\left(J\ba^\alpha\right)=\bm{0}$.  Note that the upper and lower indices cannot be switched in such expressions, as they are of different type.

Incompressibility means that the three-dimensional volume form does not change, that is, $\sqrt{\bar a}\,d_3\bar z = \sqrt{\bar a} \approx \sqrt{a}\,d_3z$ to within our approximations.
Thus, $d_3 z = \frac{h+b}{\bar b} = J^{-1}$, the inverse of the Jacobian determinant \cite{DellarSalmon05}. 
This, along with the fact that $d_t \bar b = 0$, will connect changes in column depth $h+b$ with a two-dimensional compressible flow in the plane through a continuity equation.  

In the three-dimensional problem, the fluid has uniform properties and moves incompressibly, but after integrating out the third coordinate we will obtain a two-dimensional compressible fluid, with variable properties connected to a memory of initial column depth set by the topography.

We may now write the action as an integral over the present two-dimensional area
\begin{align}
	\mathcal{A} &= \int \!\! dt \!\! \int \!\! \sqrt{a}\, d\eta^1 d\eta^2 \mathcal{L}(\eta^1,\eta^2, t \, ; \bx) \, , \nonumber \\
	\mathcal{L} &= \int_{-\bar b}^0 \!\! d_3 z\, d\eta^3 \left( \tfrac{1}{2}d_t\bx \cdot d_t\bx - gz \right) \, ,
\end{align}
eliding a three-dimensional density that would appear in every coefficient.
This action corresponds to a Saint-Venant approximation in which the vertical component of inertia is ignored in the kinetic energy, in keeping with the other small-slope approximations. No pressure-like multiplier is necessary, as only admissible variations of $\bx$ will be considered and no vertical balance laws will be derived.  
 Note that $d_3 z$ is not a function of the third coordinate and can be moved out of the inner integral.  
Integrating over $\eta^3$ yields a two-dimensional action 
\begin{align}
	\mathcal{A} &= \int \!\! dt \!\! \int \!\! \sqrt{a}\, d\eta^1 d\eta^2 (h+b)
	\left( \tfrac{1}{2}d_t\bx \cdot d_t\bx - g\left[-b + \tfrac{1}{2}(h+b)\right] \right) \, , 
\end{align}
rewriting 
$h - \tfrac{1}{2}(h+b)$ as $-b+\tfrac{1}{2}(h+b)$ in anticipation of later results.\footnote{This form would have been obtained directly by a slightly different choice of the vertical coordinate $\eta^3 = (z+b)\bar b/(h+b)$, instead of $\eta^3 = (z-h)\bar b/(h+b)$ as used here. This changes the limits of integration, and makes the lateral coordinate lines in the reference configuration share the shallow curvilinearity of the topography instead of being flat.} 
Considering variations in $\bx$ alone, using $\delta b = \nabla b \cdot \delta\bx$ and $\delta(h+b) = -(h+b)\nabla\cdot\delta\bx$ as required by incompressibility \cite{DellarSalmon05}, and noting that $d_t ( \sqrt{a}\, (h+b) ) = 0$ and $\delta ( \sqrt{a}\, (h+b) ) = 0$, we obtain the first order variation
\begin{align}
	\delta\mathcal{A} &= \int \!\! dt \!\! \int \!\! \sqrt{a}\, d\eta^1 d\eta^2 (h+b) 
	\left( -d_t^2\bx + g\left[ \nabla b - \nabla(h+b) \right] \right)\cdot\delta\bx
	+\text{boundary terms}\, . \label{EL}
\end{align}
Setting $\delta\mathcal{A} = 0$ gives rise to bulk field equations for momentum in the plane, 
\begin{align}
	 d_t^2\bx + \nabla\left(gh\right) = \bm{0} \, . \label{momentumsimplest}
\end{align}
This result is consistent with \cite{Camassa96,MilesSalmon85,DellarSalmon05, VallisBOOK, BuehlerBOOK}, but \cite{ClamondDutykh12} erroneously put the total depth $h+b$ in the gradient, which is only correct if the topography $b$ is flat. 
While the topography does not appear in the form \eqref{momentumsimplest}, it will enter through the continuity equation.

\section{Hidden structure}

This seemingly simple form of the momentum equation belies another structure.  To reveal this, first use 
\mbox{$\mathcal{A} = \int \!\! dt \!\! \int \!\! \sqrt{a}\, d\eta^1 d\eta^2 \mathcal{L} = \int \!\! dt \!\! \int \!\! \sqrt{\bar a}\, d\eta^1 d\eta^2 \mathcal{\bar L}$} to define a Lagrangian density with respect to the reference area, 
\begin{align}
	\mathcal{L} &= J^{-1}\mathcal{\bar L} = (h+b)\mathcal{\bar L }/\bar b  \, , \nonumber \\
	\mathcal{\bar L }/\bar b &=  \tfrac{1}{2}d_t\bx \cdot d_t\bx - g\left[-b + \tfrac{1}{2}(h+b)\right] \, .
\end{align}
The incompressibility constraint has led to a peculiar situation in which the referential Lagrangian density $\mathcal{\bar L}$ depends on the reference configuration $\bar b$, while the Lagrangian density $\mathcal{L}$ does not.
Next, rewrite the Euler-Lagrange equation from \eqref{EL} in a more complicated, yet illustrative form
\begin{align}
	(h+b)d_t\left(d_t\bx\right) + \nabla\left[\tfrac{1}{2}g\left(h+b\right)^2\right] = (h+b)\nabla\left(gb\right) \, , \label{momentum}
\end{align}
which could of course be simplified by dividing through to obtain
\begin{align}
	d_t\left(d_t\bx\right) + \nabla\left[g\left(h+b\right)\right] = \nabla\left(gb\right) \, , \label{momentumsimple}
\end{align}
which in turn could have been obtained more easily, either directly from \eqref{EL} or by rearranging \eqref{momentumsimplest}.
The momentum equation in any of the forms presented so far must be used in conjunction with 
\begin{align}
	d_t b &= \nabla b \cdot d_t \bx \, , \\
	d_t (h+b) &= -(h+b)\nabla\cdot d_t\bx \, , \label{continuity}
\end{align}
where the latter is just the continuity equation for the column depth.  

It is, however, easiest to relate the form \eqref{momentum} of the momentum equation to the general form discussed in \cite{SinghHanna21,*SinghHanna21correction}.  Making use of \emph{explicit} partial derivatives $\partial$, equation \eqref{momentum} can be rewritten as
\begin{align}
	J^{-1}d_t\left( \frac{\partial\mathcal{\bar L}}{\partial d_t\bx} \right) + \nabla_\alpha\left(J^{-1}  \frac{\partial\mathcal{\bar L}}{\partial \ba_\beta \bar{\ba}^\beta} \cdot \bar{\ba}^\alpha \right) = J^{-1} \frac{\partial\mathcal{\bar L}}{\partial\bx} \, ,\label{momentumgeneral}
\end{align}
using $d_t \bar b = 0$ and $\frac{\partial(h+b)}{\partial \ba_\beta \bar{\ba}^\beta} = -(h+b)\ba^\beta \bar{\ba}_\beta$, the latter being a consequence of the derivative of $J^{-1}$ with respect to the deformation gradient $\bar\nabla\bx = \ba_\beta \bar{\ba}^\beta$ \cite{GurtinFriedAnand10}. 
The inverse Jacobians appear in this and subsequent equations when writing them in terms of a present rather than a referential divergence; the latter forms will also be provided later.

The form \eqref{momentumgeneral} is certainly not obvious from the simple equation \eqref{momentumsimplest}, but it has two advantages.  It serves to isolate the explicit dependence of $\mathcal{\bar L}$ on position, which broken spatial symmetry generates the right hand source term in this balance of momentum.  And it allows us to write two additional equations, the balances of energy and pseudomomentum, in a similarly revealing way.
Following \cite{herrmannalicia1981, SinghHanna21,*SinghHanna21correction},  
these balances are respectively obtained from the projections of \eqref{momentumgeneral} onto $d_t\bx$ and $\bar\nabla\bx$, using the following separations of the total time and material derivatives in terms of explicit partials: 
\begin{align}
	d_t \mathcal{\bar L} &= \frac{\partial\mathcal{\bar L }}{\partial t} + \frac{\partial\mathcal{\bar L }}{\partial \bx}\cdot d_t\bx + \frac{\partial\mathcal{\bar L }}{\partial d_t \bx}\cdot d_td_t\bx + \frac{\partial\mathcal{\bar L }}{\partial \bar\nabla\bx} : d_t\bar\nabla\bx \, , \\
	d_\alpha \mathcal{\bar L } &= \frac{\partial\mathcal{\bar L }}{\partial \eta^\alpha} + \frac{\partial\mathcal{\bar L }}{\partial \bx}\cdot d_\alpha\bx + \frac{\partial\mathcal{\bar L }}{\partial d_t\bx}\cdot d_\alpha d_t \bx + \frac{\partial\mathcal{\bar L }}{\partial \bar\nabla\bx} : d_\alpha \bar\nabla\bx \, ,
\end{align}
where the double contractions pair legs of like type (referential or present).
Both double derivatives may be permuted, the referential basis is time-independent, and $d_\alpha \mathcal{\bar L }\,\bar{\ba}^\alpha = \bar\nabla \mathcal{\bar L } =  \bar\nabla_\alpha ( \mathcal{\bar L } \bar{\ba}^\alpha ) =  J \nabla_\alpha ( J^{-1} \mathcal{\bar L } \bar{\ba}^\alpha )$, leading to the general forms
\begin{align}
	J^{-1}d_t\left( \frac{\partial\mathcal{\bar L}}{\partial d_t\bx}\cdot d_t\bx - \mathcal{\bar L} \right) + \nabla_\alpha\left( J^{-1} \frac{\partial\mathcal{\bar L}}{\partial \ba_\beta \bar{\ba}^\beta} : d_t\bx \bar{\ba}^\alpha \right) &= -J^{-1}\frac{\partial\mathcal{\bar L}}{\partial t} \, ,  \label{energygeneral} \\
	J^{-1} d_t\left( \frac{\partial\mathcal{\bar L}}{\partial d_t\bx}\cdot  \ba_\beta \bar{\ba}^\beta \right) + \nabla_\alpha\left[ J^{-1} \bar{\ba}^\beta \left( \frac{\partial\mathcal{\bar L}}{\partial \ba_\gamma \bar{\ba}^\gamma} : \ba_\beta\bar{\ba}^\alpha - \mathcal{\bar L}\delta^\alpha_\beta\right) \right] &= -J^{-1} \frac{\partial\mathcal{\bar L}}{\partial \bar\bx}  \, . \label{pseudogeneral}
\end{align}
The source terms in these energy and pseudomomentum balances respectively arise from explicit dependences of $\mathcal{\bar L}$ on time and the reference configuration.

Identifying the Hamiltonian density and the components of a tensor formed by having the deformation gradient act on the left leg of the Eshelby tensor,
\begin{align}
	\mathcal{\bar H}/\bar b &= \left( \frac{\partial\mathcal{\bar L}}{\partial d_t\bx}\cdot d_t\bx - \mathcal{\bar L} \right)/\bar b =  \tfrac{1}{2}d_t\bx \cdot d_t\bx + g\left[-b + \tfrac{1}{2}(h+b)\right]  \, , \\
	\mathcal{\bar T}^\alpha_\beta / \bar b &=\left( \frac{\partial\mathcal{\bar L}}{\partial \ba_\gamma \bar{\ba}^\gamma} : \ba_\beta\bar{\ba}^\alpha - \mathcal{\bar L}\delta^\alpha_\beta\right) / \bar b = \left( - \tfrac{1}{2}d_t\bx \cdot d_t\bx + g\left[-b + (h+b)\right]  \right) \delta^\alpha_\beta\, ,
\end{align}
one can write the balances as
\begin{align}
	(h+b)d_t \left(\mathcal{\bar H}/ \bar b\right) + \nabla \cdot \left[ \tfrac{1}{2}g\left(h+b\right)^2 d_t \bx \right] &= 0 \, , \label{energy} \\
	(h+b)d_t\left(d_t\bx\cdot \ba_\beta \bar{\ba}^\beta\right) + \nabla_\alpha\left[ \left(h+b\right) \left(\mathcal{\bar T}^\alpha_\beta / \bar b\right) \bar{\ba}^\beta \right] &= \frac{h+b}{\bar b}\frac{\mathcal{\bar L }}{\bar b}  \bar\nabla\bar b \, , \label{pseudomomentum} 
\end{align}
recalling that $d_t \bar{\ba}^\beta = \bm{0}$.
Equations (\ref{energy}-\ref{pseudomomentum}) can also be written as follows,
\begin{align}
	(h+b)d_t \left( \tfrac{1}{2} \left[ d_t\bx \cdot d_t\bx + g(h-b) \right] \right) + \nabla \cdot \left[ \tfrac{1}{2} g\left(h+b\right)^2 d_t \bx \right] &= 0 \, , \label{energy2} \\
	(h+b)d_t\left(d_t\bx\cdot \ba_\beta \bar{\ba}^\beta \right) + \nabla_\beta\left[ \left(h+b\right) \left( - \tfrac{1}{2}d_t\bx \cdot d_t\bx + gh \right) \bar{\ba}^\beta \right] &= \frac{h+b}{\bar b} \tfrac{1}{2}\left[ d_t\bx \cdot d_t\bx - g(h-b) \right] \bar\nabla\bar b \, . \label{pseudomomentum2} 
\end{align}
Although it is not obvious from cursory inspection, equations \eqref{energy} and \eqref{pseudomomentum} can be respectively derived by projecting \eqref{momentum} onto $d_t\bx$ and $\bar\nabla\bx$ and rearranging, making use of Piola identities and other tricks.  The necessary manipulations are very unlikely to be performed without foreknowledge of some formal structure. 
Considering a time-dependent gravity, as in B{\"{u}}hler \cite{BuehlerBOOK}, should generate a source term in the energy balance instead of the conservation law \eqref{energy}.
The pseudomomentum source in \eqref{pseudomomentum} arises from non-flatness of the reference topography, and thus non-uniformity of the fluid column depth per referential area, effectively a material property of the fluid sheet.  Whitham \cite{Whitham65-2} applied Noether's theorem to derive both the energy and pseudomomentum equations in the case of flat topography.  He also derived the continuity equation from a Lagrangian. 

In the referential forms of these equations, the referential topography $\bar b$ is no longer hidden.
The momentum equation is (compare with \eqref{momentum} or \eqref{momentumsimple}),
\begin{align}
	d_t\left(\bar b d_t\bx\right) + \bar\nabla_\alpha\left[\tfrac{1}{2}\bar b g\left(h+b\right)\ba^\alpha\right] &= \bar b \nabla\left( gb \right) \, , \label{momentumref}
\end{align}
while the energy and pseudomomentum equations are (compare with (\ref{energy2}-\ref{pseudomomentum2})),
\begin{align}
	d_t \left( \tfrac{1}{2} \bar b \left[ d_t\bx \cdot d_t\bx + g(h-b) \right] \right) + \bar\nabla_\alpha \left[ \tfrac{1}{2} \bar b g\left(h+b\right) \ba^\alpha\cdot d_t \bx \right] &= 0 \, , \label{energyref} \\
	d_t\left(\bar b d_t\bx\cdot \ba_\beta \bar{\ba}^\beta \right) + \bar\nabla \left[ \bar b \left( - \tfrac{1}{2}d_t\bx \cdot d_t\bx + gh \right) \right] &=  \tfrac{1}{2}\left[ d_t\bx \cdot d_t\bx - g(h-b) \right] \bar\nabla\bar b \, , \label{pseudomomentumref} 
\end{align}
recalling that $d_t \bar b = 0$.  
Regardless of the forms chosen to represent the divergences, the momentum and pseudomomentum equations are always, respectively, present and referential vector equations.
The momentum equation \eqref{momentum} or \eqref{momentumsimple} can be easily projected onto the present basis to get equations for the components of acceleration (in the present basis), 
whereas the pseudomomentum equation \eqref{pseudomomentumref} can be easily projected onto the referential basis to get equations for the rates of change of components of momentum (again in the present basis). 
 Though related, these are different equations,
\begin{align}
	&d_t\left(d_t\bx\right)\cdot \ba_\alpha \;\,+ d_\alpha\left[g\left(h+b\right)\right] \qquad\qquad\quad= d_\alpha\left(gb\right) \, , \label{momentumproj}\\
	&d_t\left(\bar b d_t\bx\cdot \ba_\alpha \right) + d_\alpha \left[ \bar b \left( - \tfrac{1}{2}d_t\bx \cdot d_t\bx + gh \right) \right] =  \tfrac{1}{2}\left[ d_t\bx \cdot d_t\bx - g(h-b) \right] d_\alpha \bar b \, , \label{pseudomomentumproj}
\end{align}
and \eqref{momentumsimple} and \eqref{pseudomomentumref} are distinct balance laws.
For example, one difference in the effect of the source terms is that a gradient in topography $b$ can cause fluid to accelerate downhill, while a 
 gradient in the referential column depth $\bar b$ can 
  drive a 
  non-uniform velocity field that spreads material markers. 

\section{Concluding remarks} 

Looking beyond this rudimentary shallow water system, the same approach may be applied to higher-order Lagrangian-derived models such as the Green-Naghdi equations \cite{MilesSalmon85}. 
However, before doing so, one might wonder if anything has been gained from the present example, in which a 
simple equation like \eqref{momentumsimplest} was transformed into several more complicated expressions.
The approach taken is likely consistent with those applied to the pseudomomentum of disturbance flows, where the ``reference configuration'' is a base flow that may or may not be steady-state.  Now that both approaches have been applied to the same type of problem, it should be a bit easier to illuminate the connections between these ideas.
Additionally, the final comment of the previous section suggests that pseudomomentum may be a useful concept in understanding the driving force for spreading of films, in settings dominated by inertia rather than viscous forces.

\section*{Acknowledgments}

I am indebted to H. Singh for help with the machinery, as well as for many helpful arguments over the years.

\bibliographystyle{unsrt}

\end{document}